\documentclass[11pt, oneside]{article} 
\usepackage[utf8]{inputenc}
\usepackage[utf8]{inputenc}
\usepackage{geometry}                		
\usepackage{graphicx}				
\usepackage{amssymb}
\usepackage{amsmath}
\usepackage{subfig}
\usepackage{caption}
\usepackage{comment}
\usepackage{amsthm,verbatim,amsfonts,amscd,float,physics}
\usepackage{appendix}

\usepackage[utf8]{inputenc}
\usepackage{xcolor}
\numberwithin{equation}{section}
\usepackage{titling, lipsum}
\title{Infinity Wars: 
Three Types of Singularities in Non-supersymmetric Canonical Gravity and String Theory}
\author{ Michael McGuigan\\
 email contact: michael.d.mcguigan@gmail.com
}
\date{}

\begin{document}
\begin{titlingpage}

\maketitle
\begin{abstract}
We discuss three separate types of infinities that occur in non-supersymmetric canonical gravity and string theory. We consider UV perturbative singularities in non-supersymmetric gravity coupled to matter and how these are related to loop corrections to beta functions for non-supersymmetric strings. Next we consider classical cosmological singularities that occur in these equations and discuss a specific singular cosmology of Dudas and Mourad for non-supersymmetric string theory. Finally we discuss the infinities that occur in quantum cosmology associated with topology change and discuss how non-supersymmetric string quantum cosmology can be used to address them.

\end{abstract}
\end{titlingpage}

\section{Introduction}

Although infinities have been largely resolved in the standard model of particle physics there still remains several types of infinities that occur when one couples gravity to matter. This is particularly true in non-supersymmetric thories where one does not have a Bose-Fermi symmetry that can cancel divergences. Non-supersymmetric gravity plus matter  can diverge even at the one-loop level. Also a coupling of gravity to electromagnetic radiation is enough to induce a big bang type singularity in the solution to Einstein's equations. Finally there are topology changing transitions which are particular to quantum gravity and are forbidden in the classical theory and can can lead to discontinuous transitions and various types of pinching, conical and curvature singularities. There are several proposed solutions to such infinities in the context of loop quantum gravity, asymptotic safety and string theory but there is no overarching understanding of how to resolve them. The different approaches each have their virtues and drawbacks however they are not mutually exclusive. Indeed we shall see that a multitude of approaches are likely to be necessary to address the different types of singularities which have a very different nature and require a different formalism to formulate and discuss their solution. 

In this paper we discuss the three type of singularities in the context of non-supersymmetric string theory. This theory has recently been of great interest as supersymmetry has not yet been discovered at the LHC and in addition the measurement of dark energy has drawn attention to the compatiblity of de Sitter space and string theory \cite{Blaszczyk:2014qoa}
\cite{Abel:2015oxa}
\cite{Ashfaque:2015vta}
\cite{McGuigan:2019gdb}. Non-supersymmetric string theory gives us an example of a string theory with a positive cosmological constant albeit one in ten dimensions \cite{Alvarez-Gaume:1986ghj}
\cite{Dixon:1986iz}. This paper is organized as follows. In section two we discuss perturbative infinities that occur in gravity plus matter at the one loop level. We discuss how in the context of non-supersymmetric string theory one can balance the infinities that occur at the one-loop level with tree level divergences that occur in string theory in de Sitter space through the Fischler-Susskind mechanism. We also discuss the limitations of this approach at high energies beyond the black hole threshold and when non-perturbative contributions are important. In section three we discuss cosmological singularities that occur in non-supersymmetric string theory in particular focusing on the singular cosmology of Dudas and Mourad. In section four we discuss how one can use a canonical gravity formulation to address topology change in the context of non-supersymmetric string cosmology and discuss the extensions and limitations of the approach. Finally in section five we discuss the main conclusions of the paper and suggest some directions for further study.

\section{Perturbative infinities in gravity plus matter}

For some theories, like $N=8$ supergravity, one has to do complex multiloop caclulations to investigate potential divergences, but for non-supersymmetric matter coupled to gravity, only  one-loop calculations are sufficient. For example studying gravity coupled to scalar matter using dimensional regularization one finds the following divergence at one-loop \cite{tHooft:1974toh}\cite{Coquereaux:1989dm}\cite{Edwards:2022qiw}
\cite{Bastianelli:2002fv}
\cite{Bastianelli:2013tsa}
\begin{equation}\Delta L = \frac{{\sqrt { - g} }}{{8{\pi ^2}(d - 4)}}\frac{{203}}{{80}}{R^2}\end{equation}
For Einstein-Yang-Mills theory we also have a one-loop divergence given by \cite{Deser:1974nb}
\cite{Deser:1974xq}
\begin{equation}\Delta L = \frac{{\sqrt { - g} }}{{8{\pi ^2}(d - 4)}}\left\{ {\left( {\frac{{137}}{{60}} + \frac{{r - 1}}{{10}}} \right)R_{\mu \nu }^2 + {C_2}\frac{{11}}{{12}}F_{\mu \nu }^{a2}} \right\}\end{equation}
where $r$ is the dimension of the gauge group, and $rC_2=C^{abs}C_{abc}$ and $C_{abc}$ are the structure constants of the gauge group.
Finally for pure non-supersymmetric gravity at two-loops one has the divergence \cite{Goroff:1985th}
\cite{Goroff:1985sz}:
\begin{equation}\Delta L = \frac{{\sqrt { - g} }}{{{{\left( {4\pi } \right)}^4}(d - 4)}}\frac{{209}}{{2880}}R_{\alpha \beta }^{\gamma \delta }R_{\gamma \delta }^{\mu \nu }R_{\mu \nu }^{\alpha \beta }\end{equation}
As these divergences involve terms that do not appear in the original Lagrangian these theories all non-renormalizable. One can still treat these theories from an effective action point of view and discuss physical effects, however the effective Lagrangian is mainly restricted to low energies \cite{Donoghue:1994dn}
\cite{Woodard:2009ns}
\cite{Woodard:2014jba}
\cite{Carlip:2015asa}
\cite{Binetti:2022xdi}. It is better to have a UV complete theory from which one can obtain the effective gravity coupled to matter as a low energy limit. Some examples include asymptotic safety, loop quantum gravity, causal dynamical triangulations, and string theory. Most of these come with additional requirements like an ultraviolet fixed point, discrete space-time or extra dimensions. In the case of non-supersymmetric string theory one has a divergence at one-loop coming from infrared effects and a dilaton tadpole which destabilizes the vacuum. If one uses the beta function approach to string quantization one finds the following beta functions at tree level genus $g=0$ and one loop genus $g=1$\cite{Callan:1988wz}.
\begin{align}&\beta _{\mu \nu }^{(g = 0)} = {R_{\mu \nu }} + 2{\nabla _\mu }{\nabla _\nu }\phi \nonumber\\
&\beta _{\mu \nu }^{(g = 1)} = e^{2\phi}{\lambda _{10}}{g_{\mu \nu }}\end{align}
where $\lambda_{10}$ is a on-loop ten dimensional cosmological constant. These divergences look forbidding and is one reason that people did not actively pursue non-supersymmetric string theory, however these divergences can be cancelled against each other through the Fischler-Susskind mechanism \cite{Fischler:1986ci}
\cite{Fischler:1986tb}. Here one adds the divergence from one-loop to the tree level beta function to cancel the divergence. Although one has separate divergences for genus zero if the background is not Ricci flat and genus one if the one-loop cosmological constant is non zero in the end one has a background de Sitter metric with curvature determined by the value of the one-loop cosmological constant which is positive for non-supersymmetric string theory. Forming the combination of beta functions one can arrange the equations in the form:
\begin{align}&{\beta ^\phi } = \frac{1}{2}R - {\nabla ^2}\phi  - \frac{1}{2}{(\nabla \phi )^2}\nonumber\\
&{\beta _{\mu \nu }} = {R_{\mu \nu }} + 2{\nabla _\mu }{\nabla _\nu }\phi  - {\lambda _{10}}{e^{2\phi }}{g_{\mu \nu }}\end{align}
For matter stress energy tensor $T^m_{\mu\nu}$ we  recover the Einstein equations with a one loop correction coming from the cosmological constant term by combining the beta function equations to form:
\begin{equation}{R_{\mu \nu }} - \frac{1}{2}R{g_{\mu \nu }} + e^{2\phi}{\lambda _{10}}{g_{\mu \nu }}= {T^m_{\mu \nu }}\end{equation}
The Fischler-Susskind mechanism can be shown to work for arbitrary genus in the sense that if it cancells the divergence at genus $g$ one can adjust the background to cancel the divergence coming from the $g+1$ contribution \cite{La:1989kw}
\cite{Russo:1989kq}. One can still object to the Fischler-Susskind mechanism for non-supersymmetric string theory as it does not take into account non-perturbative corrections to the beta functions when curvatures become large \cite{Banks:2019oiz}
\cite{Banks:2003vp}
\cite{Banks:1998vs} or one has to take into account asymptotic darkness where high energies are dominated by the creation of black holes \cite{Basu:2010nf}
\cite{Banks:1999gd}. However in that case the interior of the black holes created has a singularity and for strong curvature we can have a big bang  singulariity. These types of  singularities are separate from perturbative infinities coming from Feynman diagrams and are considered in more detail in the next section.

\section{Classical cosmological infinities}

Classical singularities in space-time metrics are important in early Universe cosmology and in the description of the black hole interior which contains a singularity \cite{Belinski:2017fas}
\cite{Belinski:2009wj}
\cite{Belinsky:1981vdw}
\cite{Penrose:1964wq}
\cite{Landsman:2021mjt}
\cite{Landsman:2022hrn}
\cite{Senovilla:2022vlr}. Attempts to resolve these singularities have been made using loop quantum gravity \cite{Ashtekar:2008ay}
\cite{Corichi:2009pp}, asymptotic safety \cite{Adeifeoba:2018ydh} and string theory \cite{Easson:2003ia}
\cite{McAllister:2007bg}
\cite{Nappi:1992kv}
\cite{Berkooz:2002je}
\cite{Giveon:2003gb}
\cite{Lawrence:1995ct}
\cite{Elitzur:2002vw}
\cite{Russo:2003ky}
\cite{Nekrasov:2002kf}
\cite{Craps:2007ch}. Notably for certain sigma models in string theory, singularities can be either resolved or the string can propagate in the space-time with singularities as in the case of an orbifold. It is difficult to resolve these space-time singularities in general though. 

In this section we use the Einstein frame metric instead of the string frame metric used to describe the beta functions in the previous section. The metric in the string frame is related to that of the Einstein frame through:
\begin{equation}g_{\mu \nu }^{(S)} = {e^{\phi /2}}g_{\mu \nu }^{(E)}\end{equation}
The ten dimensional metric-dilaton-antisymmetric-tensor-gauge-field effective action for the non-supersymmetric $SO(16)\times SO(16)'$ heterotic string in the Einstein frame is:
\begin{equation}S_{10d} = \frac{1}{2\kappa^2}\int {{d^{10}}x} \sqrt { - g} (R - \frac{1}{2}{(\nabla \phi )^2} - 2{e^{\phi /2}}({\lambda }{e^{2\phi }} ) - \frac{1}{4}{e^{ - \phi /2}}(F_1^2 + F_2^2) - \frac{1}{{12}}{e^{ - \phi }}{H^2})\end{equation}
where $H$ is the field strength of the antisymmetric tensor and $F_1,F_2$ are the field strengths  of $SO(16$ and $SO(16)'$.
For simplicity we consider the uncompactified ten dimensional theory with metric and dilaton field. Compactifications of the non-supersymmetric string on coset manifolds, tori, orbifolds and flux compactifications are considered in \cite{Lust:1986kj}
\cite{Ginsparg:1986wr}
\cite{Nair:1986zn}
\cite{Font:2002pq}
\cite{Baykara:2022cwj}
\cite{Raucci:2022bjw}
\cite{Basile:2018irz}. Non-supersymmetric  compactifications of type II string theory are considered in \cite{Berglund:2001aj}
\cite{Berglund:2019pxr}\cite{Berglund:2021xlm} with the interesting feature of having a de Sitter space metric on a bubble of positive vacuum energy while avoiding naked singularities. Also because the exponential dilaton potential is not favored experimentally \cite{Vacher:2023gnp} we will study compactified non-supersymmetric string theory and different dilaton potentials in future work. The effective action for Einstein-dilaton theory is then:
\begin{equation}{S_{10d}} = \frac{1}{{2{\kappa ^2}}}\int {{d^{10}}x\sqrt { - g} \left[ {R - \frac{1}{2}{{\left( {\partial \phi } \right)}^2} - 2\lambda {e^{5\phi/2 }}} \right]} \end{equation}
with
$\kappa  = {{\alpha '}^2}$ and the one-loop cosmological constant given by:
\begin{equation}\lambda  = \frac{1}{{{{\alpha '}^5}}}\frac{{{2^6}}}{{{{\left( {2\pi } \right)}^{10}}}}5.67\end{equation}
We will use the ansatz:
\begin{align}
&d{s^2} =  - N{(t)^2}d{t^2} + a{(t)^2}(\sum\limits_{i = 1}^9 {d{x_i}d{x_i}} )\nonumber\\
&\phi  = \phi (t)\end{align}
and units such that $\alpha'=1$.
The Ricci scalar $R$ is given by:
\begin{equation}R = 18\frac{{\ddot a}}{{{a}{N^2}}} + 72\frac{{{{\dot a}^2}}}{{{a^2}{N^2}}}   - 18\frac{{\dot a\dot N}}{{a{N^3}}}\end{equation}
The action using this ansatz after integrating by parts is:
\begin{equation}S = \frac{1}{{2{\kappa^2}}}\int {dtN{a^9}\left[ { - 72\frac{{{{\dot a}^2}}}{{{N^2}{a^2}}} + \frac{1}{{2{N^2}}}{{\dot \phi }^2} - 2\lambda {e^{5\phi /2}}} \right]} \end{equation}
The equations of motion following from the action are given by:
\begin{align}
 - 36{{\dot a}^2} &+ \frac{1}{4}{{\dot \phi }^2}{a^2} + \lambda {e^{5\phi /2}}{a^2}{N^2} = 0\nonumber\\
 - 72\ddot a &= 252\frac{{{{\dot a}^2}}}{a} + \frac{9}{4}{{\dot \phi }^2}a - 9\lambda {e^{5\phi /2}}a{N^2} - 72\dot a\frac{{\dot N}}{N}\nonumber\\
\frac{1}{2}\ddot \phi  &=  - \frac{9}{2}\dot \phi \frac{{\dot a}}{a} - \frac{5}{2}\lambda {e^{5\phi /2}}{N^2} + \frac{1}{2}\dot \phi \frac{{\dot N}}{N}
\end{align}
We show a solution to these equations using the gauge $N=1$ and initial conditions such  $a=1$ and $\phi = -1$ at $t=0$ in figure 1.  Choosing the gauge 
$N{e^{5\phi /4}} = 1$
we have the exact cosmological solution of Dudas and Mourad\cite{Dudas:2000ff} written in the Einstein-frame given by:
\begin{align}
&N(t) = {\left( {\sin (\sqrt \lambda  t)} \right)^{ - 5/8}}{\left( {\cos (\sqrt \lambda  t)} \right)^{-5/2}}\nonumber\\
&a(t) = {\left( {\sin (\sqrt \lambda  t)} \right)^{1/24}}{\left( {\cos (\sqrt \lambda  t)} \right)^{ - 1/6}}\nonumber\\
&{e^{\phi (t)}} = {\left( {\sin (\sqrt \lambda  t)} \right)^{1/2}}{\left( {\cos (\sqrt \lambda  t)} \right)^2}
\end{align}
Similar string cosmological solutions were found in \cite{Tseytlin:1991xk}
\cite{Mueller:1989in}. The metric develops curvature singularities at $t=0$ and $t=\pi/(2\sqrt{\lambda})$ and is shown in figure 2. The singular component of the Ricci curvature is shown in figure 3. Such cosmological singularities signal the breakdown of the beta functions of perturbative string theory of the previous section. One approach to these singularities is string quantum cosmology where the values of the metric tensor and dilaton field $a(t), \phi(t)$ are replaced by wave functions $\Psi(a, \phi)$ obeying the Wheeler-de Witt equation \cite{Thebault:2022dmv}
\cite{Dabrowski:1997vp}
\cite{Gasperini:2021eri}
\cite{Gasperini:1996np}
\cite{Gasperini:2002bn}
\cite{Gasperini:1996fn}
\cite{Gasperini:2000dp}
\cite{Cavaglia:1999xr}
\cite{Cavaglia:1999ka}
\cite{Yan:2006gh}
\cite{Robles-Perez:2021rqt}. We shall study this approach in the next section related to a new type of infinity in nonsupersymmetric canonical gravity and string theory, topology change.

\begin{figure}
\centering
  \includegraphics[width = .75 \linewidth]{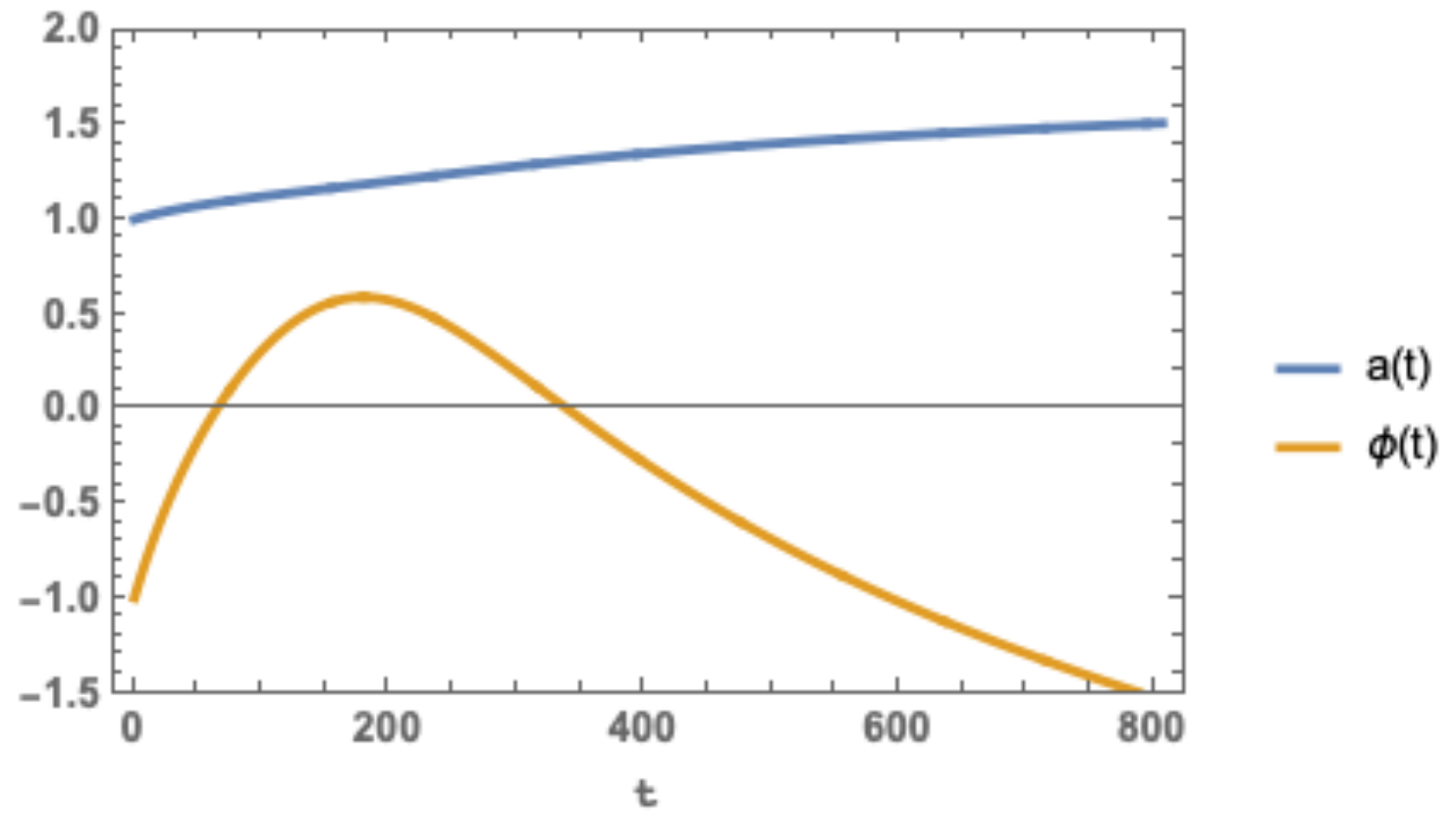}
  \caption{Scale factor and dilaton field with initial conditions such that $a_0=1$ and $\phi_0=-1$ in the gauge $N=1$. The asymmetric trajectory for the dilaton field comes from the Hubble friction term between the salce factor and dilaton field.}
  \label{fig:Radion Potential}
\end{figure}

\begin{figure}
\centering
  \includegraphics[width = .75 \linewidth]{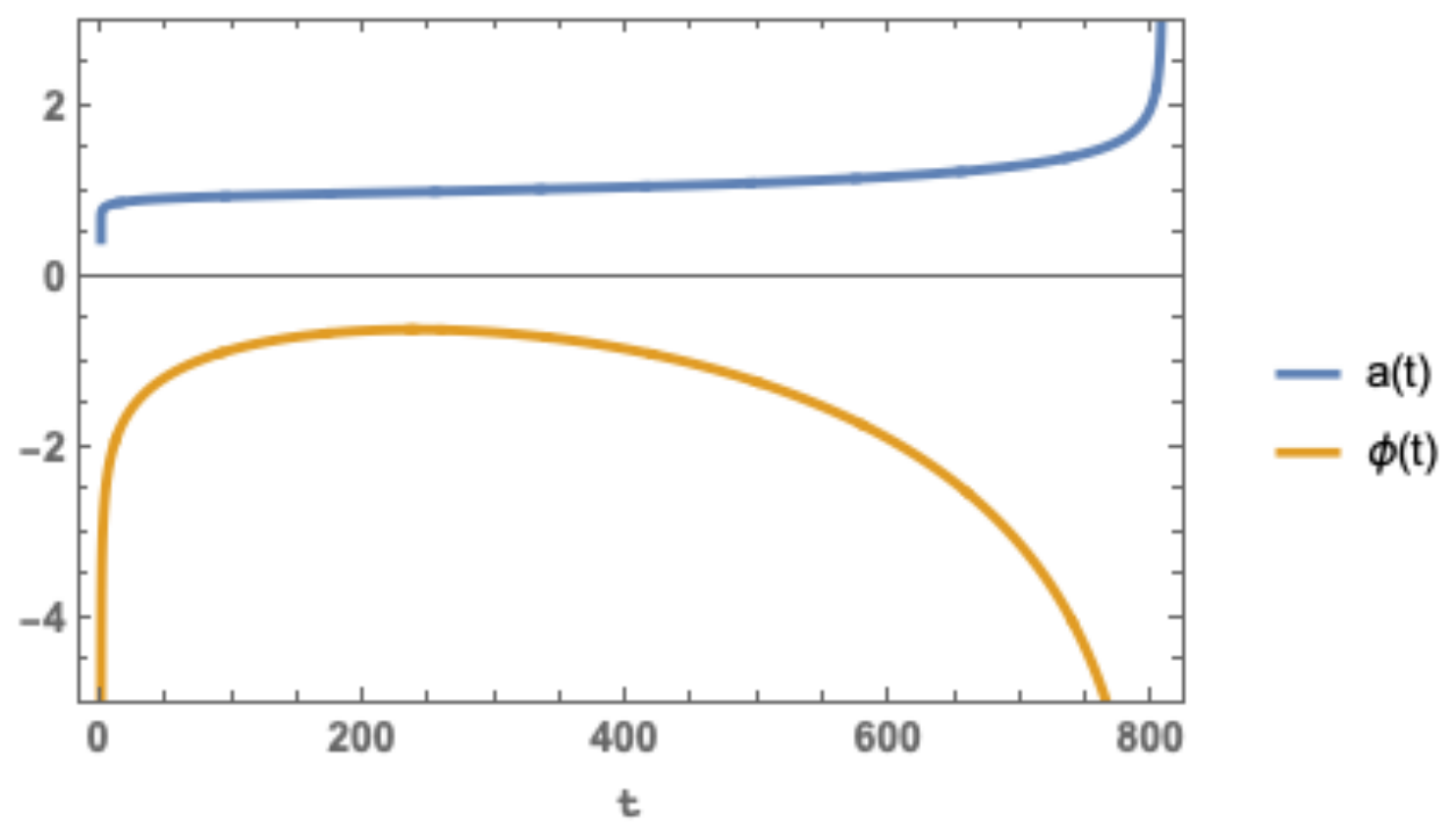}
  \caption{Scale factor and dilaton field for Duda-Mourad solution with curvature singulaties  at $t=0$ and $t=\pi /(2 \sqrt{\lambda}$) in the gauge $Ne^{5\phi/4}=1$.}
  \label{fig:Radion Potential}
\end{figure}

\begin{figure}
\centering
  \includegraphics[width = .75 \linewidth]{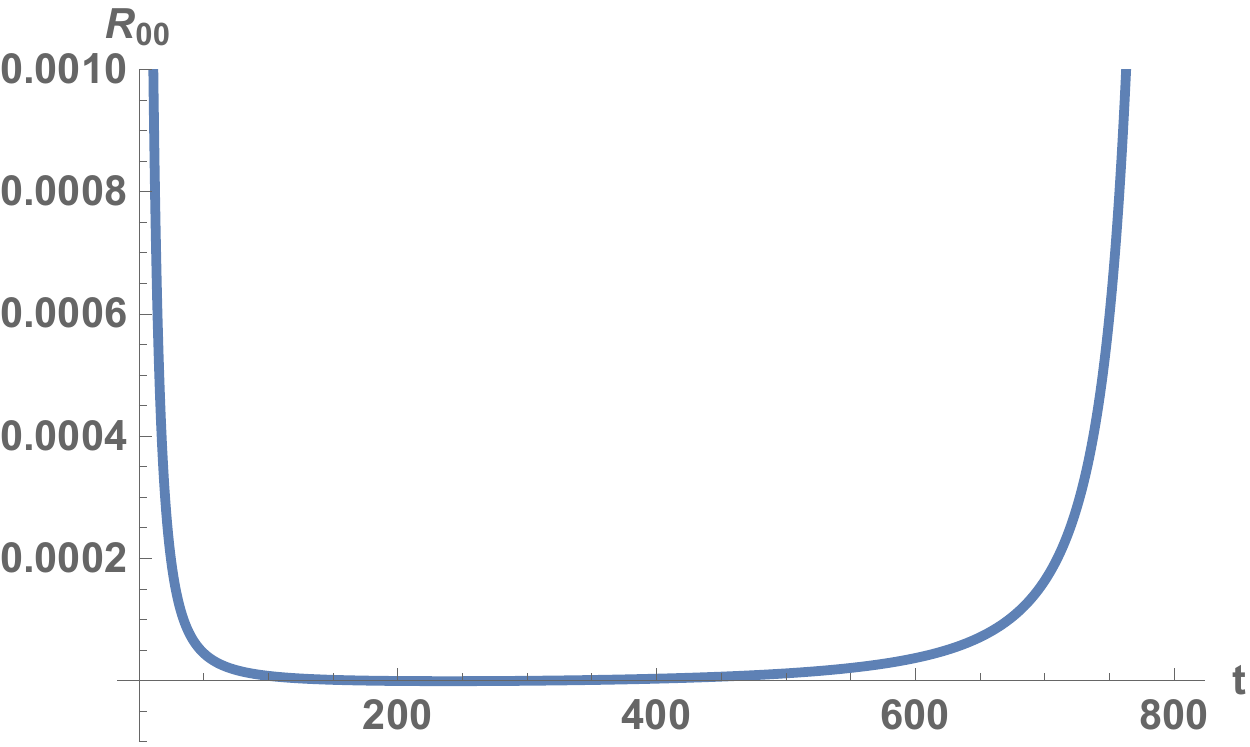}
  \caption{ $R_{00}$ component of the Ricci curvature for non-supersymmetric gravity plus dilaton theory with curvature singularities at $t=0$ and $t=\pi /(2 \sqrt{\lambda}$)}
  \label{fig:Radion Potential}
\end{figure}

\section{Quantum cosmological infinities and topology change}

The final type of infinities we wish to discuss are infinities which occur in quantum cosmology and topology change \cite{Horowitz:1990qb}
\cite{Dowker:2002hm}
\cite{Louko:1995jw}
\cite{Witten:2021nzp}
\cite{Anderson:1986ww}
\cite{Sorkin:1997gi}
\cite{Borde:1994tx}
\cite{Buck:2016ehk}
\cite{Garcia-Heveling:2022fkf}
\cite{Dowker:1999wu}
\cite{Ambjorn:2021fkp}
\cite{Greene:2000yb}
\cite{Dowker:1997kc}
\cite{Gursoy:2007np}
\cite{DeLorenci:1997hu}
\cite{Hartnoll:2003dg}
\cite{Dzhunushaliev:2009jx}
\cite{Perelman:2006up}
\cite{Lyons:1992md}
\cite{Dray}
\cite{Greene:1995hu}
\cite{Aspinwall:1993nu}
\cite{Candelas:1989ug}
\cite{Grinstein:1989pc}. Some approaches to topology change include mirror manifolds in string theory, complex Euclidean metrics, Ricci flow with surgery, Morse theory and effective vertex operators associated with the Wheeler-de Witt equation. Conical and curvature singularities can occur at points in the manifold where the topology transition occurs. A simple example of topology change is the pair of pants topology which a circle geometry transitions to a direct sum of two circles. A Lorentzian picture of this transition can lead to infinities at the point where the circle breaks into two circles. For the simple model we consider in this paper, with a two dimensional mini-superspace defined by the metric and dilaton field, we only need a vertex to describe one point splitting into two, and also a Green function to take states from one point in mini-superspace to another, similar to a world-line approach to baby Universes or quantum field theory \cite{Casali:2021ewu} .







The ten dimensional Einstein-dilaton action with boundary term is given by
\begin{equation}S_{10d} = \frac{1}{2}\int\limits_M {{d^{10}}x\sqrt { - g} } \left( {R + \frac{1}{2}{{\left( {\partial \phi } \right)}^2} - 2V(\phi )} \right) - \int\limits_{\partial M} {{d^9}x\sqrt h K} \end{equation}
with $V(\phi)$ the dilaton potential and $K$ the trace of the extrinsic curvature is defined by:
\begin{equation}{K_{ij}} = \frac{1}{{2N}}\left( {{{\dot h}_{ij}} - {D_i}{N_j} - {D_j}{N_i}} \right)\end{equation}
The canonical momentum associated with the spatial metric $h_{ij}$ is:
\begin{equation}{p^{ij}} = \sqrt h \left( {{K^{ij}} - K{h^{ij}}} \right)\end{equation}
The Hamiltonian constraint is:
\begin{equation}H = {\mathcal{G}_{ijkl}}{p^{ij}}{p^{kl}} + \frac{2}{{\sqrt h }}{p_\phi }{p_\phi } - \sqrt h \left( {{R^{(9)}} - 2V(\phi )} \right)\end{equation}
with $R^{(9)}$ the spatial curvature and the mini-superspace metric is:
\begin{equation}{\mathcal{G}_{ijkl}} = \frac{1}{{2\sqrt h }}\left( {{h_{ik}}{h_{jl}} + {h_{il}}{h_{jk}} - {h_{ij}}{h_{kl}}} \right)\end{equation}
In the variables $a,\phi$ these canonical momentum become:
\begin{align}
&{p_a} =  - \frac{1}{N}72{a^7}\dot a\nonumber\\
&{p_\phi } = \frac{1}{{2N}}{a^9}\dot \phi 
\end{align}
and the Hamiltonian constraint is:
\begin{equation}H =  - \frac{1}{{144{a^7}}}p_a^2 + \frac{1}{{{a^9}}}p_\phi ^2 + {a^9}V(\phi )\end{equation}
Defining $\alpha = \log(a)$ the Lagrangian is:
\begin{equation}L = \frac{{{e^{9\alpha }}}}{N}\left( { - 36{{\dot \alpha }^2} + \frac{{{{\dot \phi }^2}}}{4}} \right) - N\lambda {e^{9\alpha  + \frac{5}{2}\phi }}\end{equation}
Further defining:
\begin{align}
&\eta  = \frac{1}{2}\left( {5(6\alpha ) + 3\left( {\frac{\phi }{2}} \right)} \right) \nonumber\\
&\xi  = \frac{1}{2}\left( {3(6\alpha ) + 5\left( {\frac{\phi }{2}} \right)} \right)
\end{align}
the Lagrangian is written:
\begin{equation}L = \frac{{{e^{9\alpha }}}}{{4N}}\left( { - {{\dot \eta }^2} + {{\dot \xi }^2}} \right) - N\lambda {e^{9\alpha  + \frac{{5\phi }}{2}}}\end{equation}
Note in \cite{Halliwell:1986ja}
\cite{Garay:1990re} a different choice of variables was chosen to treat the exponential potential in four dimensions. Redefining $N = \bar N{e^{ - 9\alpha  - \frac{{5\phi }}{2}}}$ we have:
\begin{equation}L = \frac{{{e^{18\alpha  + \frac{{5\phi }}{2}}}}}{{4\bar N}}\left( { - {{\dot \eta }^2} + {{\dot \xi }^2}} \right) - \bar N\lambda \end{equation}
or more simply:
\begin{equation}L = \frac{{{e^{2\xi }}}}{{4\bar N}}\left( { - {{\dot \eta }^2} + {{\dot \xi }^2}} \right) - \bar N\lambda \end{equation}
The mini-superspace metric is:
\begin{equation}\delta {s^2} = {\mathcal{G}_{AB}}d{X^A}d{X^B} = {e^{2\xi }}\left( { - d{\eta ^2} + d{\xi ^2}} \right)\end{equation}
and inverse:
\begin{equation}{\mathcal{G}^{AB}} = {e^{ - 2\xi }}\left( {\begin{array}{*{20}{c}}
{ - 1}&0\\
0&1
\end{array}} \right)\end{equation}
and the Hamiltonian constraint is:
\begin{equation}H = {e^{ - 2\xi }}\left( { - p_\eta ^2 + p_\xi ^2} \right) + \lambda  = 0\end{equation}
The Wheeler-de Witt equation is:
\begin{equation}H\Psi \left( {\eta ,\xi } \right) = \left[ {{e^{ - 2\xi }}\left( {  \partial _\eta ^2 - \partial _\xi ^2} \right) + \lambda } \right]\Psi \left( {\eta ,\xi } \right) = 0\end{equation}
The mini-superspace metric is identical to the metric for $1+1$ dimensional Rindler space with solutions to the Wheeler-de Witt equation are of the form:
\begin{equation}\Psi \left( {\eta ,\xi } \right) = {e^{ - i\nu \eta }}{\psi _\nu }(\xi )\end{equation}
with
\begin{equation}{\psi _\nu }(\xi ) = {\left( {\frac{{2\nu \sinh \nu \pi }}{{{\pi ^2}}}} \right)^{1/2}}{K_{i\nu }}(\sqrt{\lambda} {e^\xi })\end{equation}
These wave functions are normalized using the Klein-Gordon inner product \cite{DeWitt:1967yk}:
\begin{equation}({\Psi _1},{\Psi _2}) =  - i\int_{ - \infty }^\infty  {d\xi \Psi _1^ * {{\mathord{\buildrel{\lower3pt\hbox{$\scriptscriptstyle\leftrightarrow$}} 
\over \partial } }_\eta }{\Psi _2}} \end{equation}
Other choices for inner product are considered in \cite{Witten:2022xxp}
\cite{Halliwell:1992cj}
\cite{Woodard:1989ac}
\cite{Carlip:1990kp}. Another quantity to compute is the Green function which is the path integral from one nine dimensional spatial slice to another. This can be computed from Euclidean or Lorentzian path integral techniques or from the product of solutions to the Wheeler-de Witt equation \cite{Halliwell:1988ik}
\cite{Brown:1990iv}
\cite{Feldbrugge:2017kzv}. Using the latter technique for the $\eta, \xi$ minisuperspace the Green function is given by \cite{Haba:2007ay}\cite{Svaiter:1989pq}:
\begin{equation}G(\eta ,\xi ;\eta ',\xi ') = \int_0^\infty  {d\nu } \frac{1}{{{\pi ^2}}}{e^{ - i\nu  {\left( {\eta  - \eta '} \right)} }}\sinh \pi \nu {K_{i\nu }}(\sqrt{\lambda }{e^\xi }){K_{i\nu }}(\sqrt{\lambda} {e^{\xi '}})\end{equation}
Returning  to $(a, \phi)$ variables we have:
\begin{align}
&{e^\eta } = {a^{15}}{e^{3\phi /4}}\nonumber\\
&{e^\xi } = {a^9}{e^{5\phi /4}}\end{align}
So the solution to the Wheeler-de Witt equation in the $(a, \phi)$ variables is:
\begin{equation}\Psi (a,\phi ) = {({a^{15}})^{ - i\nu }}{e^{ - i\nu 3\phi /4}}{K_{i\nu }}(\sqrt{\lambda} {a^9}{e^{5\phi /4}})\end{equation}
Finally variables that transform the minisuperspace metric to the Minkowski metric are:
\begin{align}
&T = {e^\xi }\sinh \eta \nonumber \\
&X = {e^\xi }\cosh \eta \end{align}
In terms of the $a,\phi$ variables these are:
\begin{align}
&T = {a^9}{e^{5\phi /4}}\frac{1}{2}\left( {{a^{15}}{e^{3\phi /4}} - {a^{ - 15}}{e^{ - 3\phi /4}}} \right)\nonumber \\
&X = {a^9}{e^{5\phi /4}}\frac{1}{2}\left( {{a^{15}}{e^{3\phi /4}} + {a^{ - 15}}{e^{ - 3\phi /4}}} \right)\end{align}

For one Universe splitting into two, an  approach is to introduce vertex operators into the path integral, similar to string theory in Rindler space. 
The vertex operator for the Rindler mini-superspace metric can be written as a sum over vertex operators for the flat Minkowski mini-superspace metric as \cite{Chamblin:2006xd}
\begin{align}
&{V_\nu } = \sqrt {\sinh \pi \nu } \frac{1}{{\pi }}{e^{ - i\omega \eta }}{K_{i\nu }}\left( {\sqrt{\lambda} {e^\xi }} \right) = 
\nonumber \\
&\frac{1}{{4\pi \sqrt {\sinh \pi \nu } }}\int_{ - \infty }^\infty  {\frac{{d\ell {e^{i\ell X}}}}{{\sqrt {\lambda  + {\ell ^2}} }}} \left\{ {{e^{\frac{{\pi \nu }}{2} - i\nu {{\sinh }^{ - 1}}\left( {\ell /\sqrt {\lambda  + {\ell ^2}} } \right) - i\sqrt {\lambda  + {\ell ^2}} T}} - {e^{ - \frac{{\pi \nu }}{2} + i\nu {{\sinh }^{ - 1}}\left( {\ell /\sqrt {\lambda  + {\ell ^2}} } \right) + i\sqrt {\lambda  + {\ell ^2}} T + }}} \right\}\end{align}
One can use the Green function in a similar manner to Green functions in a background gravitational field with the transformation between states $\Psi_i$ and $\Psi_j$ given by \cite{Chitre:1977ip}
\cite{Charach:1981hf}:
\begin{equation}{A_{ij}} =  - \int {d{\Sigma ^A}(\eta ,\xi )\int {d{\Sigma ^B}(\eta ',\xi ')} \Psi _i^ * } (\eta ,\xi ){{\mathord{\buildrel{\lower3pt\hbox{$\scriptscriptstyle\leftrightarrow$}} 
\over \partial } }_A}{\Psi _j}(\eta ',\xi '){{\mathord{\buildrel{\lower3pt\hbox{$\scriptscriptstyle\leftrightarrow$}} 
\over \partial '} }_B}G(\eta ,\xi ;\eta ',\xi ')\end{equation}
This yields an quantum approach to cosmological singularities. For example one could take initial states as a wave packet which follows a trajectory through minsuperspace of the  cosmological solution in section 3 and study its evolution to a different point in mini-superspace.
For the wave function we have the integral representation:
\begin{equation}{K_{i\nu }}\left( {\sqrt \lambda  {e^\xi }} \right) = \int_0^\infty  {\exp \left[ { - \sqrt \lambda  {e^\xi }\cosh w} \right]} \left( {\cos \nu w} \right)dw\end{equation}
For $\xi  \to  - \infty $ we have
\begin{equation}{K_{i\nu }}\left( {\sqrt \lambda  {e^\xi }} \right) \approx \frac{1}{2}{\left( {\frac{{\sqrt \lambda  }}{2}} \right)^{i\nu }}\Gamma \left( { - i\nu } \right){e^{i\nu \xi }}\end{equation}
and as $\xi  \to   \infty $
\begin{equation}{K_{i\nu }}\left( {\sqrt \lambda  {e^\xi }} \right) \approx \sqrt {\frac{\pi }{{2\sqrt \lambda  }}} \exp \left[ { - \sqrt \lambda  {e^\xi } - \frac{\xi }{2}} \right]\end{equation}
The wave function is plotted in figure 4. One has an oscillatory behavior in the region near the singularity for large negative $\xi$ and small $a$ with exponentially damped behavior for large positive $\xi$ and large $a$. 
\begin{figure}
\centering
  \includegraphics[width = .75 \linewidth]{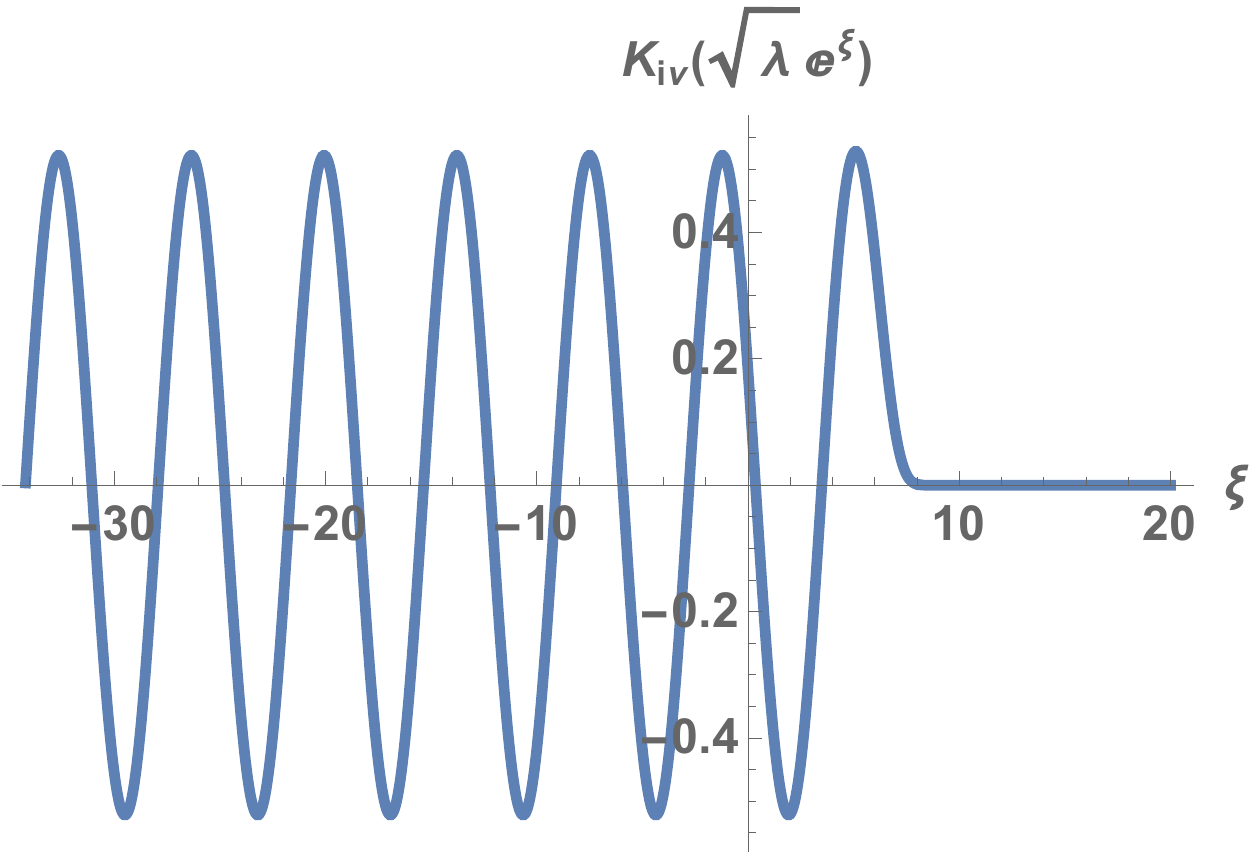}
  \caption{Quantum wave function ${K_{i\nu }}\left( {\sqrt \lambda  {e^\xi }} \right)$ for non-supersymmetric gravity-dilaton theory with $\nu =1$. The wave function shows oscillatory behavior as one approaches the singularity at small radius and exponential falloff at large radius. }
  \label{fig:Radion Potential}
\end{figure}

\section{Conclusion}

In this paper we have described three types of infinities one encounters in non-supersymmetric canonical gravity and string theory. For perturbative infinities from graviton scattering  there are several approaches. We presented one approach to  perturbative infinites involving non-supersymmetric string theory and the Fischler-Susskind mechanism. When the curvature becomes large one encounters the second type of infinity, the cosmological singularity where the scale factor $a(t)$ becomes small. The third type of singularity is associated with discontinuous transition of spacetime itself associated with topology change. This is a particularly quantum phenomenon. One approach this type of infinity is quantum cosmology where one is concerned with the wave function $\Psi(a)$ instead of $a(t)$ in the classical theory. We illustrated this in defining Universal Green functions and vertex operators for non-supersymmetric string cosmology in the mini-superspace approximation. It will be interesting to investigate going beyond the mini-superspace approximation, compactifications to four dimensions and more realistic dilaton potentials in future research. As these three types of infinities represent inconsistency and incompleteness of non-supersymmetric gravity plus matter theory, we can learn a great deal if we are able to resolve them.

\section*{Acknowledgements}
We would like to to thank Tristan Hubsch for references and correspondence about non-supersymmetric string compactifications and singularities.

\end{document}